\documentclass[conference]{IEEEtran}
\IEEEoverridecommandlockouts

\usepackage{soul}
\usepackage{booktabs}
\usepackage{multirow}
\usepackage{float}
\usepackage{url}
\usepackage[small,it]{caption}
\usepackage{tcolorbox}
\usepackage{cite}
\usepackage{amsmath,amssymb,amsfonts}
\usepackage{algorithmic}
\usepackage{graphicx}
\usepackage{textcomp}
\usepackage{xcolor}
\usepackage{url}
\def\BibTeX{{\rm B\kern-.05em{\sc i\kern-.025em b}\kern-.08em
    T\kern-.1667em\lower.7ex\hbox{E}\kern-.125emX}}
\begin{document}

\newcommand{\anthony}[1]{\textcolor{blue}{{\it [Anthony says: #1]}}}
\newcommand{\christian}[1]{\textcolor{purple}{{\it [Christian says: #1]}}}

\title{\huge Towards a Model to Appraise and Suggest Identifier Names}

\author{\IEEEauthorblockN{Anthony Peruma}
\IEEEauthorblockA{
Rochester Institute of Technology, Rochester, New York, USA \\
axp6201@rit.edu \\
\small \textit{Advisor: Christian D. Newman (cnewman@se.rit.edu), Rochester Institute of Technology, USA}
}
}

\maketitle

\begin{abstract}
Unknowingly, identifiers in the source code of a software system play a vital role in determining the quality of the system.  Ambiguous and confusing identifier names lead developers to not only misunderstand the behavior of the code but also increases comprehension time and thereby causes a loss in productivity. Even though correcting poor names through rename operations is a viable option for solving this problem, renaming itself is an act of rework and is not immune to defect injection. 

In this study, we aim to understand the motivations that drive developers to name and rename identifiers and the decisions they make in determining the name. Using our results, we propose the development of a linguistic model that determines identifier names based on the behavior of the identifier. As a prerequisite to constructing the model, we conduct multiple studies to determine the features that should feed into the model. In this paper, we discuss findings from our completed studies and justify the continuation of research on this topic through further studies.
\end{abstract}

\begin{IEEEkeywords}
Program Comprehension, Identifier Names
\end{IEEEkeywords}

\begin{comment}
\begin{itemize}
\item \href{https://goo.gl/iijtVK}{RPA Guidelines: https://goo.gl/iijtVK}
\item \href{https://goo.gl/7rL6y2}{RPA Rubric: https://goo.gl/7rL6y2}
\end{itemize}
\end{comment}

%%%%%%%%%%%%%%%%%%%%%%%%%%%%%%%%%%%%%%%%%%%%%%%%%%%%%%%%%%%%%%%%%%%
%%%%%%%%%%%%%%%%%%%%%%%%%%%%%%%%%%%%%%%%%%%%%%%%%%%%%%%%%%%%%%%%%%%

\section{Introduction}
\label{section:Introduction}
Software maintenance is the most costly phase of the software development lifecycle \cite{5726933,846201}, with a significant portion of this (about 58\%) dedicated to source code comprehension \cite{7997917}. Developers must use identifier names to comprehend the code that they will update.  Identifiers are names (i.e., lexical tokens) that uniquely identify entities in the code (such as classes, methods, variables, etc.). It has been estimated that identifiers contribute to, on average, 70\% of a software system's codebase \cite{1421019}. It has also been shown that poor identifier names can cause developers to spend, on average, 19\% more time on comprehension activities \cite{7884623}; a result supported by \cite{Lawrie2007}. It has also been repeatedly shown that the name of a good identifier explicitly reflects its role \cite{10.1007/978-3-642-03013-0_14,Liblit2006CognitivePO}. 

The need for strong identifier names is reflected in standard software engineering practices, which provide guidelines \cite{CDocs,GoogleGuide}, best practices \cite{martin2009clean,goodliffe2007code}, and quality metrics \cite{Scalabrino2018,5332232,Posnett:2011:SMS:1985441.1985454} to assist developers in naming identifiers. The idea is that,
through the use of unambiguous and intent-revealing names, identifiers assist in communicating the purpose and behavior of the source code and eventually the functionality of the system to developers, which research has shown to be true \cite{Butler:2009:EIN:1595782.1595796}.

Unfortunately, naming conventions and best practices can only guide developers to strong names; they cannot be used to provide a developer with a high-quality name, and they cannot be used to provide a holistic comparison of multiple candidate names for an identifier. Further, quality metrics for readability \cite{Scalabrino2018,5332232,Posnett:2011:SMS:1985441.1985454} do not work at the level of identifiers names; they cannot inform a developer if a name is high-quality. Instead, they explicitly look at source code structure and use static analysis to estimate, for example, complexity and use that to measure comprehension. In short, there are currently no methods that can be used to determine whether an identifier name is high-quality or not. Furthermore, there are no models that accurately tell us how developers create an identifier name that reflects source code behavior.

The goal of this work is to begin the creation of a model that understands the relationship between the name of an identifier and the behavior of the source code entity it represents. To accomplish this, we need first to understand how developers choose names for identifiers. We can achieve this by studying instances where developers rename existing identifiers in the source code (i.e., rename refactorings) %\footnote{Refactoring is an activity where the developer updates the structure of the source code without impacting the external behavior of the system \cite{fowler2018refactoring}}). 
Developers perform renames to reflect better the meaning of the identifier, which can either result in a change or preservation in meaning \cite{Arnaoudova:2014:RAN:2693206.2693288}.

To automate the process of identifier name evaluation, the features that impact an identifiers name will be used as input to our model. This will theoretically allow us to provide developers with real-time, context-aware suggestions and appraisals of identifier names. Our proposed model will reduce the time and costs involved in software maintenance and ensure that production-ready code is readable and understandable before deployment. We also envision that the broader impact of this work will drive further research into program comprehension and result in improvements to software engineering tools, for example, in the area of source code generation.% and system documentation. %Tools (such as test case generation tools) will be better equipped to generate identifiers with high-quality intent revealing names. To further support maintenance activities, there exists the possibility of extending our model using natural language techniques to auto-generate system documentation that meets human readability and understandability standards.}

%Our long term goal is to model the relationship between the name of an identifier and its behavior. However, to achieve this goal, we first need to study the renaming refactorings performed by developers. Through the study of rename refactorings, we gain an understanding of how the semantic meaning of an identifier changes and the motivations that drive developers to perform a rename.% and the consequences on the code (if any) due to the rename.
%\christian{Strong goal statement at the bottom}

%Deriving answers to these and other similar questions not only provides us with the underlying cause(s) for renames, but also provokes more research into mechanisms to assist developers from either implementing code that is less subject to identifier rename operations or correcting invalid names earlier on in the development lifecycle. Outcomes from these studies would lead to improvements in developer productivity, application quality, and project costs due to shorter developer ramp-up times and less source code rework.

%%%%%%%%%%%%%%%%%%%%%%%%%%%%%%%%%%%%%%%%%%%%%%%%%%%%%%%%%%%%%%%%%%%
%%%%%%%%%%%%%%%%%%%%%%%%%%%%%%%%%%%%%%%%%%%%%%%%%%%%%%%%%%%%%%%%%%%
\section{Rename Taxonomy}
%In software engineering, identifiers are lexical tokens that uniquely identify entities in the source code (e.g., classes, methods, variables, etc). Developers perform rename operations on these identifiers to reflect the change in the behavior of the system, correct wrongly named identifiers or ensure consistency in identifier naming.

When renaming an identifier, a developer may add, remove or replace terms to/from the original name. Either one of these actions (or even a combination of these actions) updates semantic meaning of the name. In other words, this act of renaming can either change the meaning of the name or preserve the original meaning. In this study, we utilize the taxonomy created by Arnaoudova et al. \cite{Arnaoudova:2014:RAN:2693206.2693288} to examine rename refactorings and categorize them into the different types prescribed by this taxonomy.

At its basic form, preservation of meaning occurs if the terms in the name are reordered or special characters are included/excluded. For more complex forms, meaning preservation is maintained if the replaced terms are synonyms or are a singular/plural of the original. An example of preservation in
meaning occurred when a developer renamed \textit{pictureLock} $\rightarrow$ \textit{photoLock}. In this instance, the term `picture' is a synonym of `photo' and hence preserves the meaning of the original name.

A change in meaning can occur for multiple reasons. Developers can perform a specialization of an identifiers name by replacing a term in the original name with a hyponym or adding an adjective or noun to the original name as in the instance when the developer performed the following rename: \textit{button} $\rightarrow$ \textit{customMediaRouteButton}. In this example, the newly added terms are nouns hence the semantic change is considered a narrowing in meaning. By replacing a term with a hypernym or through the removal of terms (e.g., \textit{author\_name} $\rightarrow$ \textit{contact}), a developer can generalize (i.e., broaden) the meaning of the identifiers name. Adding new terms to the identifiers name (without causing a specialization) like \textit{scanPortsButton} $\rightarrow$ \textit{scanWellKnownPortsButton}, leads to an addition to the meaning of the name. In this example, the newly added terms are an adverb and verb. Hence the semantic change is not considered a specialization, therefore it is categorized under add meaning. Similarly, the removal of terms (without causing a generalization) removes details from the meanings name (e.g.,  \textit{mPendingDeletedMessages} $\rightarrow$  \textit{mPendingMessages}). Additionally, an identifiers meaning is also modified if terms in the original name are replaced with antonyms (e.g., \textit{chartTop} $\rightarrow$ \textit{chartBottom}).

%%%%%%%%%%%%%%%%%%%%%%%%%%%%%%%%%%%%%%%%%%%%%%%%%%%%%%%%%%%%%%%%%%%
%%%%%%%%%%%%%%%%%%%%%%%%%%%%%%%%%%%%%%%%%%%%%%%%%%%%%%%%%%%%%%%%%%%
\section{Motivating Example}
A prerequisite to constructing the linguistic model is determining how and why identifier names change.  %Hence, to achieve these goals, we first study the semantic changes a name undergoes. Next, we apply our findings to determine what drives developers to perform renames by studying the co-occurrence of renames with other refactorings and correlating maintenance activities with renames.
To this extent, a key feature that we can exploit from renames is that we can see the actions performed by developers before and after a rename. In other words, using a static analysis approach, we can determine the types of refactorings that occur either before, with or after a rename refactoring. Furthermore, combining messages in the commit log and the type of semantic change that the identifier name undergoes, we can aim to contextualize the rationale behind the rename. For example, in \cite{code_example01} a developer moves a class 
from one package to another with the message: ``Incremental changes, some package refactorings etc''. The next refactoring operation \cite{code_example02} on this class is the renaming of the class from  \textit{JsonViewResult}$\rightarrow$\textit{JsonView}. This rename broadens the meaning of the name by removing the term \textit{`Result'}, making the identifier more general in meaning. The commit message for this rename is:  ``Cleaned up some file names for easier usage...'', meaning the developer was likely going through and renaming things after the move class refactoring. Considering such patterns in the implementation lifecycle of the system, there exists the possibility of extracting appropriate features from the source code to construct our proposed identifier name appraisal and recommendation model. We envision our model being available as an extension in the developers integrated development environment (IDE) thereby providing the developer with real-time suggestions and appraisals for identifier names during the implementation and maintenance phases.

%%%%%%%%%%%%%%%%%%%%%%%%%%%%%%%%%%%%%%%%%%%%%%%%%%%%%%%%%%%%%%%%%%%
%%%%%%%%%%%%%%%%%%%%%%%%%%%%%%%%%%%%%%%%%%%%%%%%%%%%%%%%%%%%%%%%%%%
\section{Approach}
Our goal of constructing a high performing linguistic model is composed of multiple studies. These are studies that aim to determine the features that are most appropriate to feed into the model. Hence, in this section, we discuss the results of our completed studies and propose future studies on this topic. %\christian{It needs to be clear how the things in subsection A feed into your proposed work. Why is it important that you've done this stuff?}

\subsection{Completed Studies}
%\vspace{2mm} \noindent \textbf{\textit{Android refactorings:}}
%We present a summary of our study of refactorings in 1,028 Android applications (apps) \cite{Peruma:2018:MobileSoft}. The basis of this study was to evaluate the feasibility of using Android apps as a dataset to study refactorings. Our findings showed that approximately 41.93\% of the mined refactorings are renames. Additionally, when compared to non-mobile Java systems, we observed differences in the distribution of refactoring types. Finally, contextualizing these refactorings revealed that developers refactor their code to improve comprehension, resolve defects, and due to functionality changes. Even though we observed a high prevalence of rename refactorings, due to the nature of mobile apps, the size and therefore refactorings were limited. On average, each app consisted of 93.16 commits and 47.79 refactorings. These values fall outside of values of non-mobile systems. Hence, Android apps are more suitable for Android specific studies and would probably require a different analysis approach. \christian{revise, or since you're short on space, maybe just remove it}

\vspace{0.5mm} \noindent \textbf{\textit{Contextualizing renames to commit messages:}}
We briefly report on the findings of our prior study \cite{Peruma:2018:EIW:3242163.3242169} in which we investigated the contextualizing of different semantic changes of identifier names. %More specifically, we aimed to answer the research question: \textit{``To what extent can commit message be used to contextualize different types of semantic change rename refactorings?''}
A topic modeling analysis using Latent Dirichlet Allocation (LDA) yielded interesting results but proved insufficient to pinpoint the developer's intention. Words like \textit{ad} and \textit{add} frequently occurred with narrow, broaden and add meaning categories. This may indicate that the addition of code correlates with these types of changes. Interestingly, we observed that preserve and remove meaning lacked the words \textit{ad} and \textit{add} in their commit message.  If we assume adding tends to modify meaning somehow (i.e., narrow, broaden, add) then preserve and remove meaning should not include these terms. Instead, terms like \textit{rename} and \textit{refactor} are more common in these categories than in others.  Interestingly, remove meaning does not include terms like \textit{remove}, \textit{delete}, etc. Even though commit messages provide interesting trends around semantic updates to identifier names, they cannot be solely relied on for contextualizing renames. In other words, results from our LDA analysis did not yield clear-cut topics for each of the semantic categories. There were overlapping of terms between some topics, or some topics were missing key terms that are usually associated with the semantic change. The topics generated by LDA were too high-level; unable to provide us with fine-grain data about the context around renaming practices. Therefore, in the next study, we considered co-occurrence of other refactorings with renames.

\vspace{1.5mm} \noindent \textbf{\textit{Co-occurrence of renames and other refactorings:}} This study \cite{scam2019} involved identifying refactorings preceding or following a rename. Our study of 800 systems showed that in most scenarios, renaming of an element does not generally seem to be influenced by, nor does itself influence another type of refactoring on the same element. However, there is a subset of renames that occur directly before or after another refactoring. From this subset, we observed that a majority of the time developers perform a refactoring operation just before the rename, these two operations happen in a short (commit) interval. We also showed that developers frequently change the semantic meaning of an identifier name when performing a rename after a refactoring. Contextualizing these refactorings with the commit log proved useful for filtering out a set of commit messages closely related to different types of renames. However, while the rationale for some semantic changes can be derived from the commit log in addition to the refactorings that occurred just prior to the rename, we still encountered high-level LDA topics. In other words, the level of detail described by developers about their activities/tasks in the commit message is not sufficient for fine-grained NLP-based analysis. This made it difficult to understand the reasoning behind the application of renames fully. Hence, our findings show that a significant amount of work is needed to automatically derive these motivations more effectively from commit messages, other natural language software artifacts, and general source code changes. A final contribution from this study showed that developers with limited project experience are more inclined to perform only rename refactorings than other types of refactorings (that may alter the systems design); indicating that renames are applied by developers that may not be very familiar with the system they are developing for.

\vspace{1.5mm} \noindent \textbf{\textit{Abbreviations in source code:}} In addition to our work on renames, we also investigated abbreviation expansion. The act of expanding abbreviations is valuable for studying identifiers since it allows us to remove the threat that a developer might not be familiar with a given abbreviation, and makes it easier for tools such as part of speech taggers to work effectively. In \cite{ICSME_2019}, we studied the expansion of abbreviations that appear in the source code of five open-source systems. This study enabled us to understand the effectiveness of different abbreviation expansion techniques on systems with varying quality of documentation. Additionally, we manually created a gold set of over 850 abbreviation-expansion pairs. This will help us study the effect of abbreviations on comprehension and has the potential to increase the quality of our proposed model.

%\hl{Because the results from our analysis of renames, commits, and refactorings were too high-level to serve as data for project-specific rename context, the next step is to begin investigating renames from the perspective of developers. That is, we will study how developers apply renames and what information they use to create a name in their heads. Additionally, we will talk to developers after they apply a rename to understand their personal reasoning for changing each word in the target identifier. This data will assist us in determining what data sources are important to consider when building our model. It will also help us determine ways to automate ways to contextualize renames, since we will know what developers tend to look for when applying renames.} \christian{Clean this up a little, but this is how I'd think about transitioning. You can combine this with your paragraph below}

\vspace{2mm} In summary, while results from our analysis of identifier renames, commits, and refactorings were promising, these results do not provide a complete picture on the motivations that drive developers to rename identifiers. In short, the results were too high-level to serve as data for project-specific rename contexts. Hence, to construct our model, we need to extend our investigation to look beyond static analysis and begin investigating renames from the perspective of developers. That is, we will study how developers apply renames and their thought process in determining a name. Additionally, we would need to interact with developers to understand their reasoning for changing each word in the target identifier. This data will assist us in determining what data sources are important to consider when building our model. It will also help us determine ways to automate the contextualization of renames since we will know what developers tend to look for when applying renames.

\subsection{Proposed Studies}
As explained in our completed studies, we have shown that even though we did notice patterns in identifier renames, trying to contextualizing these renames using static analysis of source code is not straightforward. The results we obtained are very generic, and, at most, provides more of a high-level outline as to why developers perform rename operations. While useful, these results cannot be directly incorporated into our proposed model. Hence, we need to further investigate developer implementation (and maintenance) activities to derive features for our model. As such, we need to shift away from empirical studies and focus on studies where we have explicit developer involvement. With developer involvement, we will be able to derive a more fine-grained rationale behind identifier renames, and also the thought process involved in deciding on a new, and more appropriate, name. Therefore, going forward, we propose a more identifier-oriented exploratory study on the developer's viewpoint of identifier renames. This exploratory study will constitute of an eye-tracking study on developer actions and reactions to identifier names in source code. From this study, we aim to answer the following research questions:

{\small
\vspace{1.5mm} \noindent
\textbf{RQ1}: What source code elements do developers look at when renaming an identifier?

\vspace{1mm} \noindent
\textbf{RQ2}: Do developers look at certain code elements more when applying different types of semantic changes?

%\vspace{1mm} \noindent
%\textbf{RQ3}: To what extent is the relationship between the time spent by a developer looking at an identifier name and the semantic change applied to the identifier after its renamed? 

\vspace{1mm} \noindent
\textbf{RQ3}: What are the trends in the types of semantic change applied to an identifier and the reason a developer applied that semantic change?
}
% and 2) a case study on well engineered and popular software systems. %\christian{Make this stronger. You have a set of problems that -need- a solution and your efforts so far leave you with a gap in knowledge. What is that gap? And then in the next paragraph, you argue why eye tracking and a case study are the appropriate ways to fill this gap}

\vspace{1.5mm}
Eye-tracking in software engineering studies is not new. It is primarily utilized for studies that involve the comprehension of software artifacts such as models and source code \cite{SHARAFI201579}. Other studies have used this technology to study developer interactions in performing change tasks \cite{Kevic:2015:TSD:2786805.2786864}, defect identification \cite{Sharif:2012:ESR:2168556.2168642}, and debugging \cite{7312518} among others. However, at present, there does not exist work that focuses on software refactorings, and more specifically rename refactorings. With eye-tracking, the medium of our study will be the developer's environment, and we plan on utilizing iTrace \cite{Guarnera:2018:IET:3204493.3208343} to integrate eye-tracking into this environment. This will provide us with the opportunity to determine elements or concepts in the source code (or even in the IDE) that developers rely on when either performing a rename or comprehending a rename. Unlike in static analysis, where we are presented with after-the-fact results associated with a rename, eye-tracking provides us with the ability to capture/measure concepts such as fixations, scanpaths, and areas of interest \cite{7467288} which are, in reality, part of the implementation process. With these concepts, we can refine the efficiency of our proposed model. For example, while static analysis informs us of the refactorings that occur prior to a rename, eye-tracking will aid us in understanding the number and types of elements developers refer to before performing a rename. Additionally, this technique can act as a proxy to measure the comprehensibility of identifier names and its likelihood to undergo a rename. For instance, a developer renaming an identifier after a relatively high gaze/fixture duration can act as an indicator of a poor name. Furthermore, patterns around the semantic change a name undergo based on the gaze/fixture duration can be used to support developers in their naming activities. Finally, we plan on interviewing the participants of the study to understand their rationale for performing renames (if any) and their thought process on deciding the new name for the identifier. As a means of mitigating risks involved in the experiment, we plan on conducting a series of trials to uncover shortcomings in our methodology (such as task clarity/complexity, participant behavior, and environmental factors). Additionally, each participant in the experiment will be allotted time to become familiar with the environment before the commencement of the experiment.

\section{Literature Review}
\label{section:LiteratureReview}
We divided our discussion of related work into two areas - studies that explore identifier renamings from a natural language perspective and studies that investigate the quality attributes that an identifier should exhibit.

\subsection{Identifier Renaming}
A survey on identifier renaming conducted by Arnaoudova et al. \cite{Arnaoudova:2014:RAN:2693206.2693288} showed that developers primarily perform renames in conjunction with other refactorings with most of the renamings due to updates in existing functionality. In the same study, the authors proposed REPENT, an approach to first detect identifier renamings in the source code, and then analyze and classify the detected renamings based on their semantic change. Through an empirical study, the authors demonstrate a high accuracy of their approach in the detection of identifier renamings and show the impact of proper naming has on minimizing software development effort.

Allamanis et al.~\cite{Allamanis:2014:LNC:2635868.2635883} implemented NATURALIZE, a framework that utilized statistical language models in mining natural source code naming conventions. The authors demonstrated the high accuracy of NATURALIZE by utilizing it in a sample set of open source projects. %Primarily proposed as a tool to supplement release management or code reviews, 
NATURALIZE learns the coding conventions in the source code via syntactic restrictions, sub-grammars on existing identifier names and utilizes this knowledgebase to evaluate snippets of new code for potential variables that should be renamed. In an extension of their work \cite{Allamanis:2015:SAM:2786805.2786849},  the authors proposed an approach to suggest renaming methods based on their bodies and renaming classes based on their methods. Their recommendation uses a neural probabilistic language model to input a set of words %, coded in a vector, 
which is fed to a hidden layer of the neural network. %The output layer evaluates the conditional probability of each word in the vocabulary given the input sequence.

Liu et al. \cite{7097720} proposed a simplistic approach to identify renaming opportunities due to renaming refactorings. Their approach relies upon the source code containing identifiers that are similar to the renamed identifier. If a similar identifier is detected, their approach recommends the developer to rename these identifiers as well. An empirical study on four applications yielded high precision values for their approach. An exploratory study on the lexical similarities between method arguments and parameters by Liu et al. \cite{7886980} demonstrated mixed results, i.e., either very low or high similarities. The authors state that further research is needed.

Studies by H{\o}st and {\O}stvold \cite{10.1007/978-3-642-00434-6_20,10.1007/978-3-642-03013-0_14} on method names showed that even though method names and behavior are mutually dependent, there is more research required in this area to better determine high-quality names. 

In summary, prior work in this area has focused on either the type of semantic change occurring in a renamed identifier or the identification of identifiers that are candidates for renaming based on similar identifiers in the codebase. However, research into the thought process of a developer in determining the rationale for performing a rename or, for that matter, determining the correct choice of a replacement name is lacking.

\subsection{Identifier Quality}
In terms of metrics for source code readability and understandability measurements, most research has focused on complete code snippets \cite{Posnett:2011:SMS:1985441.1985454,5332232}. However, within these quality models are components that focus on identifier name characteristics such as length, number of dictionary terms the identifier comprises of, and the broadness/specialization of the name. Studies on the length of identifier names by Lawrie et al. \cite{Lawrie2007} and Hofmeister et al. \cite{7884623} show that names consisting of abbreviations are harder to comprehend than full-word identifiers. Similarly, a study by Schankin et al. \cite{Schankin:2018:DCI:3196321.3196332} shows that descriptive names improve program comprehension. Studies on systems using camel case and underscores for identifier names \cite{5090039, 5521745} have shown that developer experience plays an important part in comprehending such names. %Butler et al. \cite{5328661,5714430} showed statistically significant results between poor identifier names and code quality issues.

Arnaoudova et al. \cite{Arnaoudova2016} proposed a catalog of linguistic anti-patterns in the source code. The seventeen anti-patterns span across methods and attributes. Contained in this catalog are anti-patterns that are related to the name of the identifier and its purpose/behavior. Furthermore, via a developer survey, the authors confirmed that the presence of linguistic anti-patterns in source code is a poor programming practice. Using a subset of these linguistic anti-patterns Fakhoury et al. \cite{Fakhoury:2018:EPS:3196321.3196347} demonstrated that a developers' cognitive load increases when reviewing code containing such anti-patterns.

As described, current research has focused on the quality of code snippets/chunks and not on individual identifier names. While this corpus of studies considers structural characteristics of an identifiers name, they do not consider the relationship between the name of the identifier and its intended purpose nor do they provide a formal definition for high-quality names.

%%%%%%%%%%%%%%%%%%%%%%%%%%%%%%%%%%%%%%%%%%%%%%%%%%%%%%%%%%%%%%%%%%%
%%%%%%%%%%%%%%%%%%%%%%%%%%%%%%%%%%%%%%%%%%%%%%%%%%%%%%%%%%%%%%%%%%%
%\begin{comment}
\section{Challenges and Constraints}
\label{section:Challenges}
%As with any research study, we too have encountered challenges in conducting this study. 
This section highlights only the key challenges and constraints we encountered. Our research is constrained to Java as the external tools used in our studies are Java specific. Additionally, obtaining a representative dataset is also challenging; we are constrained to open-source Java systems and these systems vary vastly in size. %As all prior work is also based on Java systems, switching programming languages will impact comparison (and replication) activities. 
Presently, there does not exist a goldset of high-quality identifier names for us to study and use as benchmarks. Similarly, a software engineering specific set of stopwords are not available for text prepossessing activities, which is prerequisite for topic modeling and n-gram analysis.
%\end{comment}
%%%%%%%%%%%%%%%%%%%%%%%%%%%%%%%%%%%%%%%%%%%%%%%%%%%%%%%%%%%%%%%%%%%
%%%%%%%%%%%%%%%%%%%%%%%%%%%%%%%%%%%%%%%%%%%%%%%%%%%%%%%%%%%%%%%%%%%

\section{Conclusion}
\label{section:Conclusion}
Identifiers play an essential role in informing developers about the behavior of the software system. Poor identifier names result in increased code comprehension time and hence, loss in developer productivity. To address this issue of poor names, developers perform renaming operations on identifiers. However, renames are considered rework and can hurt code quality and developer productivity. To help developers name identifiers with high-quality names during implementation, we need to understand the thought process of the developer. Our static analysis based research has shown us that contextualizing semantic changes of identifier names with commit messages is not sufficient. Therefore, we propose studies to investigate concepts that cannot be captured through static code analysis. Hence, our future studies will involve eye-tracking. We envision our findings feeding into a linguistic model that provides developers with real-time, context-aware identifier name suggestions during implementation.

%%%%%%%%%%%%%%%%%%%%%%%%%%%%%%%%%%%%%%%%%%%%%%%%%%%%%%%%%%%%%%%%%%%
%%%%%%%%%%%%%%%%%%%%%%%%%%%%%%%%%%%%%%%%%%%%%%%%%%%%%%%%%%%%%%%%%%%

\bibliographystyle{ieeetr}
\bibliography{references,ref-intro}

%%%%%%%%%%%%%%%%%%%%%%%%%%%%%%%%%%%%%%%%%%%%%%%%%%%%%%%%%%%%%%%%%%%
%%%%%%%%%%%%%%%%%%%%%%%%%%%%%%%%%%%%%%%%%%%%%%%%%%%%%%%%%%%%%%%%%%%

\end{document}